\documentclass[aps,prl,twocolumn,superscriptaddress,longbibliography]{revtex4-1}

\usepackage{graphicx}
\usepackage{bm}
\usepackage{hyperref}
\usepackage{todonotes}
\usepackage{color}
\usepackage{comment}
\usepackage{verbatim}
\usepackage{soul}

\begin{document}

\title{Possible quantum paramagnetism in compressed Sr$_2$IrO$_4$}

\author{D. Haskel}
\email{haskel@anl.gov}
\affiliation{Advanced Photon Source, Argonne National Laboratory, Argonne, Illinois 60439, USA}

\author{G. Fabbris}
\email{gfabbris@anl.gov}
\affiliation{Advanced Photon Source, Argonne National Laboratory, Argonne, Illinois 60439, USA}

\author{J. H. Kim}
\affiliation{Advanced Photon Source, Argonne National Laboratory, Argonne, Illinois 60439, USA}

\author{L. S. I. Veiga}
\affiliation{Advanced Photon Source, Argonne National Laboratory, Argonne, Illinois 60439, USA}
\affiliation{Brazilian Synchrotron Light Laboratory (LNLS), Campinas, SP 13083-970, Brazil}
\affiliation{Instituto de F\'isica “Gleb Wataghin”, Universidade Estadual de Campinas, Campinas, SP 13083-859, Brazil}

\author{J. R. L. Mardegan}
\affiliation{Advanced Photon Source, Argonne National Laboratory, Argonne, Illinois 60439, USA}
\affiliation{Instituto de F\'isica “Gleb Wataghin”, Universidade Estadual de Campinas, Campinas, SP 13083-859, Brazil}

\author{C. A. Escanhoela Jr.}
\affiliation{Advanced Photon Source, Argonne National Laboratory, Argonne, Illinois 60439, USA}
\affiliation{Brazilian Synchrotron Light Laboratory (LNLS), Campinas, SP 13083-970, Brazil}

\author{S. Chikara}
\affiliation{Advanced Photon Source, Argonne National Laboratory, Argonne, Illinois 60439, USA}

\author{V. Struzhkin}
\affiliation{Geophysical Laboratory, Carnegie Institution of Washington, Washington, DC 20015, USA}

\author{T. Senthil}
\affiliation{Department of Physics, Massachusetts Institute of Technology, Cambridge, Massachusetts 02139, USA}

\author{B. J. Kim}
\affiliation{Max-Planck-Institut f\"ur Festk\"orperforschung, Heisenbergstr. 1, Stuttgart D-70569, Germany}
\affiliation{Department of Physics, Pohang University of Science and Technology, Pohang 790-784, Republic of Korea}

\author{G. Cao}
\affiliation{Department of Physics, University of Colorado, Boulder, Colorado 80309, USA}

\author{J. W. Kim}
\affiliation{Advanced Photon Source, Argonne National Laboratory, Argonne, Illinois 60439, USA}

\date{\today}

\begin{abstract}
\ \ \ The effect of compression on the magnetic ground state of Sr$_2$IrO$_4$ is studied with x-ray resonant techniques in the diamond anvil cell. The weak interlayer exchange coupling between square-planar 2D IrO$_2$ layers is readily modified upon compression, with a crossover between magnetic structures around 7 GPa mimicking the effect of an applied magnetic field at ambient pressure. Higher pressures drive an order-disorder magnetic phase transition with no magnetic order detected above 17-20 GPa. The persistence of strong exchange interactions between $\mathrm{J_{eff}}=1/2$ magnetic moments within the insulating IrO$_2$ layers up to at least 35 GPa points to a highly frustrated magnetic state in compressed Sr$_2$IrO$_4$ opening the door for realization of novel quantum paramagnetic phases driven by extended $5d$ orbitals with entangled spin and orbital degrees of freedom.

\end{abstract}

\maketitle

The sizable spin-orbit interaction acting on $5d$ electrons of heavy transition metal (TM) ions together with the large spatial extent of $5d$ orbitals and related reduction/enhancement of Coloumb/crystal-field interactions leads to the emergence of novel exotic ground states in some of their oxide forms, chief amongst them tetravalent iridium compounds with half-filled $5d$ bands and $J_{\rm eff}=\frac{1}{2}$ states \cite{Nat-Phys-NaIO, Pyrochlore, Balents, Modic, Kim-Science,Kim2008, 214-SC-1, 214-SC-2, Senthil, Shitade, Jackeli, Chapon, Arita,Bertinshaw2019}. Most notably, bond-directional exchange anisotropy arising from spin-orbit coupling is expected to enhance frustration in honeycomb \cite{Nat-Phys-NaIO, Takayama-PRL} and triangular/kagome lattices \cite{Kimchi, Shimizu, Takayama-kagome} leading to novel quantum spin liquid (QSL) ground states \cite{Balents-QSL} such as the one predicted by Kitaev \cite{Kitaev}. The square lattice of Sr$_2$IrO$_4$ (Sr-214) is also capable of harboring frustration and quantum paramagnetic phases, e.g., within the $J_1-J_2-J_3$ model \cite{Figueirido1989, liquid2, Ferrer1993, Mambrini2006, Reuther2011, Danu2016} if second and/or third neighbor exchange interactions become a sizable fraction of, and compete with, first neighbor exchange interactions. The connection between QSLs and superconductivity in square lattices, via the resonating valence bond model of Anderson \cite{Anderson}, raises prospects that new forms of superconductivity could be found in iridates \cite{Senthil}. In fact, the single layer Sr-214 is nearly isostructural and displays the same spectrum of magnetic excitations as the La$_2$CuO$_4$ parent compound of high $T_c$ cuprates \cite{Kim-magnons,Liu-NatPhys} and recent doping experiments reinforce this connection \cite{214-SC-1,214-SC-2}.

Since magnetically ordered and quantum paramagnetic phases (such as spin liquids \cite{Anderson,liquid1,liquid2,liquid3} and valence bond solids \cite{vbs}) compete for spin-spin correlations in space and time, it seems desirable to continuously tune competing exchange interactions with pressure in order to suppress magnetic order, without the need for chemical substitutions and unwanted lattice disorder. Pressure has recently been used to drive honeycomb iridates away from magnetically ordered states \cite{Takayama-PRL, muSR-beta, Breznay, Veiga-I}, although observation of dimerized phases driven by formation of Ir$_2$ molecular orbitals across edge-shared IrO$_6$ octahedra \cite{Dimer-I,Dimer-II,Veiga-II} highlights the complexity and diversity of accessible magnetic ground states in compressed lattices. The rigid network of corner-shared IrO$_6$ octahedra underlying the square lattice of Sr-214 is robust against Ir dimerization and provides a suitable platform for continuously tuning exchange interactions while preserving crystal structure.  

\begin{figure*}
\includegraphics[width = 17.5 cm]{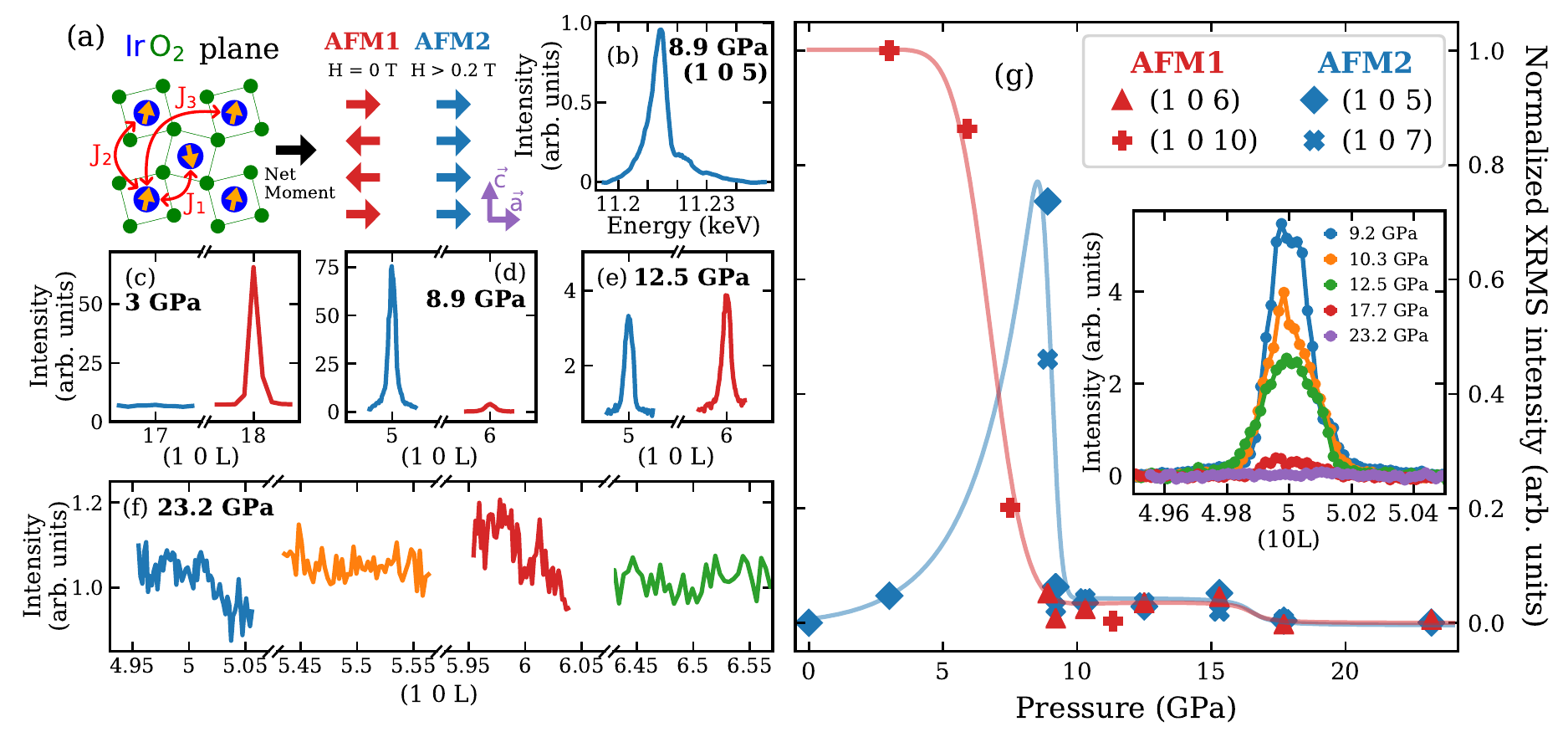}
\caption{(a) Schematic of the in-plane magnetic arrangement in Sr-214 at ambient pressure. Iridium (oxygen) ions are shown in blue (green). $J_{1,2,3}$ denote  first, second, and third neighbor exchange constants. The $c$-axis stacking of net magnetic moments in $\rm{IrO_2}$ layers in zero field (AFM1 order) and applied field (AFM2 order)\cite{Kim-Science} is also shown. (b) Resonance enhancement of the (1 0 5) magnetic peak across the Ir $L_3$ resonance ($2p_{3/2}\rightarrow 5d$, E=11.215 keV) for P=8.9 GPa, T= 10 K. (c-f) Pressure-dependent single crystal  resonant diffraction from selected (1 0 L) magnetic reflections measured in transmission (Laue) geometry at T=10 K in zero field (Neon pressure medium). Magnetic intensities are normalized to those of (1 1 10) lattice peak. Peak count rates at magnetic (1 0 5), (1 0 6) and lattice (1 1 10) peaks for P=8.9 GPa, T=10 K are 2.4$\times 10^5$ photons/sec, 7.2$\times 10^3$ photons/sec, and 2.8$\times 10^8$ photons/sec, respectively. Reflections in (c-e) panels showed typical resonant enhancement as in (b) while no enhancement was observed in residual background shown in (f). (g) Integrated intensities of selected magnetic peaks normalized to those of nearby (1 1 10) lattice peak. The solid lines are guides. The inset shows evolution of (1 0 5) integrated intensity at selected pressures (T=10 K). A crossover between magnetic structures takes place at P$\sim$7 GPa, followed by coexistence and continuous suppression of magnetic order with no detectable magnetic scattering at $\sim$ 23 GPa, as seen in this panel's inset.}
\label{fig1}
\end{figure*}

In this Letter we report on the use of pressure to tune exchange interactions in Sr-214 and drive it into what appears to be a quantum paramagnetic state. Using x-ray resonant magnetic scattering (XRMS) and x-ray magnetic circular dichroism (XMCD) measurements in the diamond anvil cell we were able to probe the evolution of both the antiferromagnetic (AFM) structure in reciprocal space (in zero applied field) and the weak ferromagnetic (W-FM) response in applied field across a critical pressure $P_c\sim$ 17-20 GPa where magnetic order vanishes. XMCD measurements of the magnetic susceptibility in the putative quantum paramagnetic phase indicate that strong local AFM exchange interactions remain present to at least 35 GPa with a Curie-Weiss temperature $\theta_{\rm CW}=-209(40)$ K, despite the lack of discernible magnetic order down to T=1.6 K even in large applied field H=6 T, indicative of a high degree of frustration within the IrO$_2$ 2D layers. The exciting prospect of a quantum critical point separating N\'eel order and quantum paramagnetic phases in compressed Sr-214 underpins the need for theoretical and experimental efforts aimed at understanding how frustration of magnetic interactions emerges in square lattices with extended $5d$ orbitals and entangled spin and orbital degrees of freedom.

The magnetic structure of Sr$_2$IrO$_4$ at ambient pressure has been determined by both x-ray \cite{Kim-Science} and neutron \cite{Ye} diffraction. The predominant Ir-O-Ir superexchange interaction ($J_1 = 60$ meV \cite{Kim-magnons}) combined with the Dzyaloshinkii-Moriya (D-M) interaction \cite{DM1,DM2} drive a canted antiferromagnetic arrangement within the IrO$_2$ layer [Fig.~\ref{fig1}(a)]. The ground state magnetic structure features an alternating stacking pattern along the c-axis [AFM1, Fig.~\ref{fig1}(a)] giving rise to $(1, 0, 4n+2)$ magnetic reflections \cite{Kim-Science}. A modest applied field ($>$ 0.2 T) is sufficient to modify the weak interlayer coupling (0.6 meV \cite{Calder2018}) and give rise to a Zeeman-energy-driven ferromagnetic stacking sequence of net moments along the c-axis [AFM2, Fig.~\ref{fig1}(a)], $(1, 0, 2n+1)$ magnetic reflections, and weak ferromagnetism $0.05-0.075 \mu_B$/Ir \cite{Kim-Science,Cao}.  

\begin{figure}
\includegraphics[width = 8.5 cm]{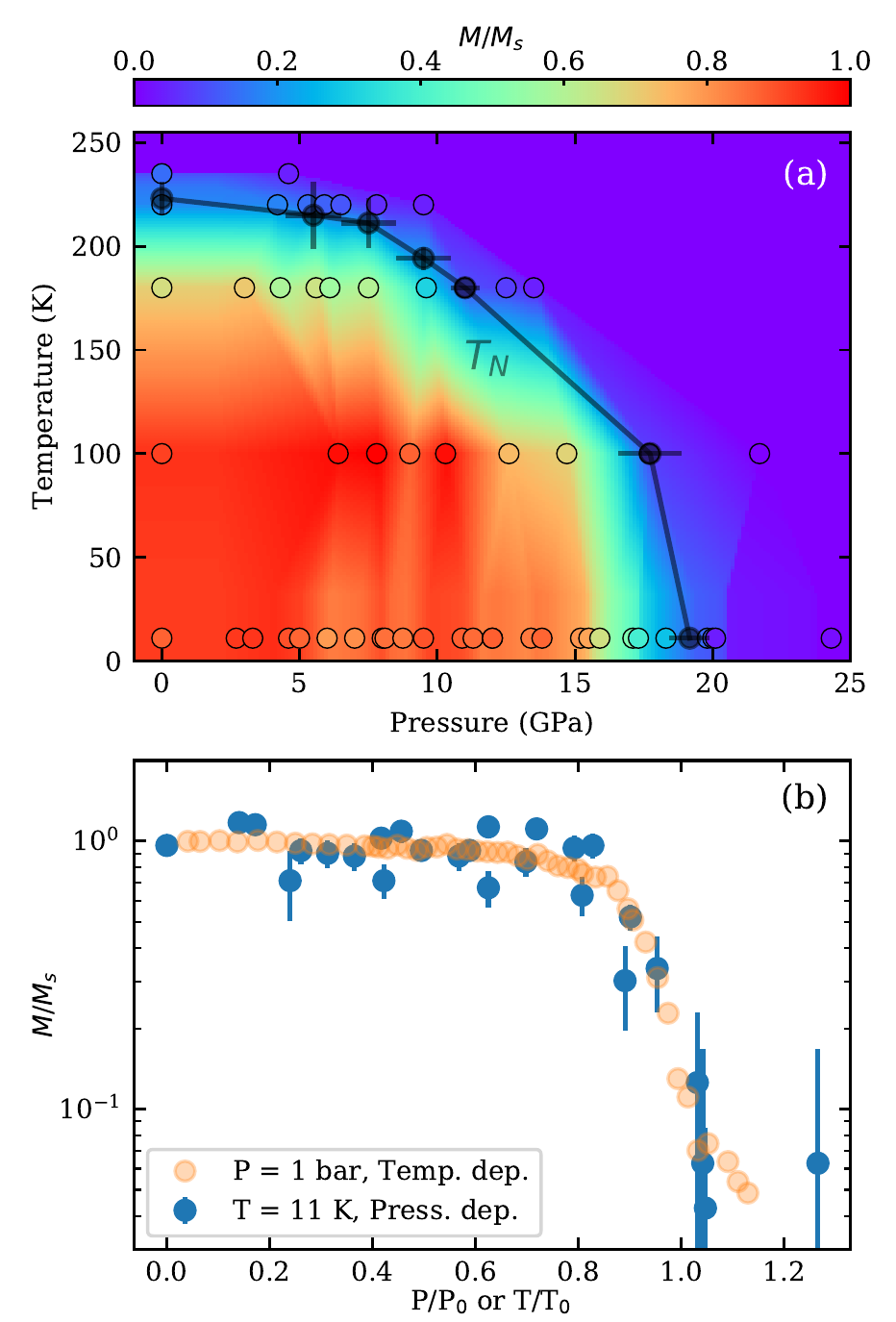}
\caption{(a) Pressure-temperature evolution of Ir magnetization in H=0.5 T applied field for a powder sample of Sr$_2$IrO$_4$ (ground single crystal) in Neon pressure medium. Circles denote P/T values at which XMCD data were collected along isotherms. Color scale represents net magnetization normalized to saturation M$_s$(P=1 bar, T=5.6 K). Unity corresponds to 3\% XMCD signal and net moment 0.05 $\mu_B$/Ir. The N\'eel temperature T$_N$ at selected pressures, shown by solid black circles, was obtained by constraining M(T) to its functional form at ambient pressure. (b) Magnetization plotted as function of reduced pressure P/P$_0$ (T=11 K, 0.5 T, P$_0$=19.2 GPa) and reduced temperature T/T$_0$ (P=1 bar, 0.8 T, T$_0$=250 K).}
\label{fig2}
\end{figure}

The evolution of the magnetic structure with pressure probed by XRMS in zero applied field and T=10 K is shown in Figs.~\ref{fig1}(c)-(g). Details on the experimental setup are found in the Supplemental material \cite{Supplemental}. A dramatic crossover between AFM1 and AFM2 magnetic structures takes place around 7(1) GPa [Fig. \ref{fig1}(g)]. Pressure, then, at first mimics the effect of an applied field, a modest c-axis compression $\sim 0.9\%$ at 7 GPa \cite{Haskel-PRL,Granado-XRD} having a dramatic effect on the interlayer exchange interactions. While the AFM2 phase is stabilized in a narrow pressure range around 8 GPa at the expense of the AFM1 phase, a strong suppression of the XRMS intensity from either phase is observed above $\sim9$ GPa ($\times$ 25 intensity reduction at 10 GPa). Further pressure causes the weak XRMS intensities from coexisting phases to vanish at about 18 GPa [Fig.~\ref{fig1}(f)]. The absence of magnetic Bragg peaks at L and L~$+\frac{1}{2}$ for odd and even L values points to the suppression of magnetic order. In particular, this observation negates the presence of collinear, in-plane AFM order as such state would lead to $(1,0, {\rm even})$ magnetic peaks as observed in Ba$_2$IrO$_4$ \cite{Boseggia, Moser}. This conclusion is also supported by the persistent IrO$_6$ rotations at high pressures \cite{Supplemental}. Despite possible evidence for an increased tetragonal distortion of IrO$_6$ octahedra \cite{Haskel-PRL, Granado-XRD}, the absence of $(1,0,{\rm odd})$ magnetic peaks also discards a spin-flop transition into a c-axis collinear AFM structure \cite{Jackeli}, as seen in Mn-doped Sr-214 \cite{Calder-Mn}.

Despite the strong in-plane exchange interaction, 3D magnetic order in Sr$_2$IrO$_4$ is stabilized at ambient pressure by the weak interlayer coupling. Thus the observed collapse of 3D magnetic order at high pressures could be attributed to a frustrated interlayer coupling. However, in such case one would expect an applied magnetic field to lift such frustration and drive magnetic order. To address this possibility, we constructed an Ir $L_3$ XMCD (P,T) phase diagram by collecting data in applied field for selected isotherms [Fig.~\ref{fig2}(a)]. A modest magnetic field H=0.5 T stabilizes the AFM2 phase at low pressures and temperatures, leading to a measurable XMCD signal that is proportional to the Ir magnetization \cite{Haskel-PRL}. The pressure range between 10 and 15 GPa is particularly noteworthy [P/P$_0$ = 0.52-0.78 in Fig.~\ref{fig2}(b), P$_0$=19.2 GPa], since a large drop of the zero field XRMS intensity in this pressure range [Fig.~\ref{fig1}(g)] is contrasted by a small reduction of the H=0.5 T XMCD signal ($\lesssim 30\%$), emphasizing the ability of a magnetic field to drive 3D magnetic order in Sr-214 even in the presence of frustrated interlayer coupling. Therefore, the combined collapse of XMCD and XRMS signals beyond 18-20 GPa is compelling evidence that such magnetic transition fundamentally involves the magnetism and/or magnetic interactions within the IrO$_2$ layers.  The evolution of magnetization with pressure at 11 K plotted versus reduced pressure, P/P$_0$, maps to that of the temperature induced order-disorder transition at 1 bar [Fig.~\ref{fig2}(b), T$_0$=250 K] pointing to second-order character which is also seen in a gradual reduction of $T_N$ with pressure [Fig.~\ref{fig2}(a)]. This behavior is consistent with evolution towards a quantum critical point but inconsistent with a transition into incommensurate spiral or other unconventional phases with different magnetic symmetry, expected to be first order. Furthermore, the consistent critical pressures P$_0$ for single crystalline and polycrystalline samples, together with lattice strain values below about 0.1\% at 23 GPa~\cite{Supplemental}, rule out lattice disorder as the primary driver for the magnetic transition.

\begin{figure}[t]
\includegraphics[width = 8.5 cm]{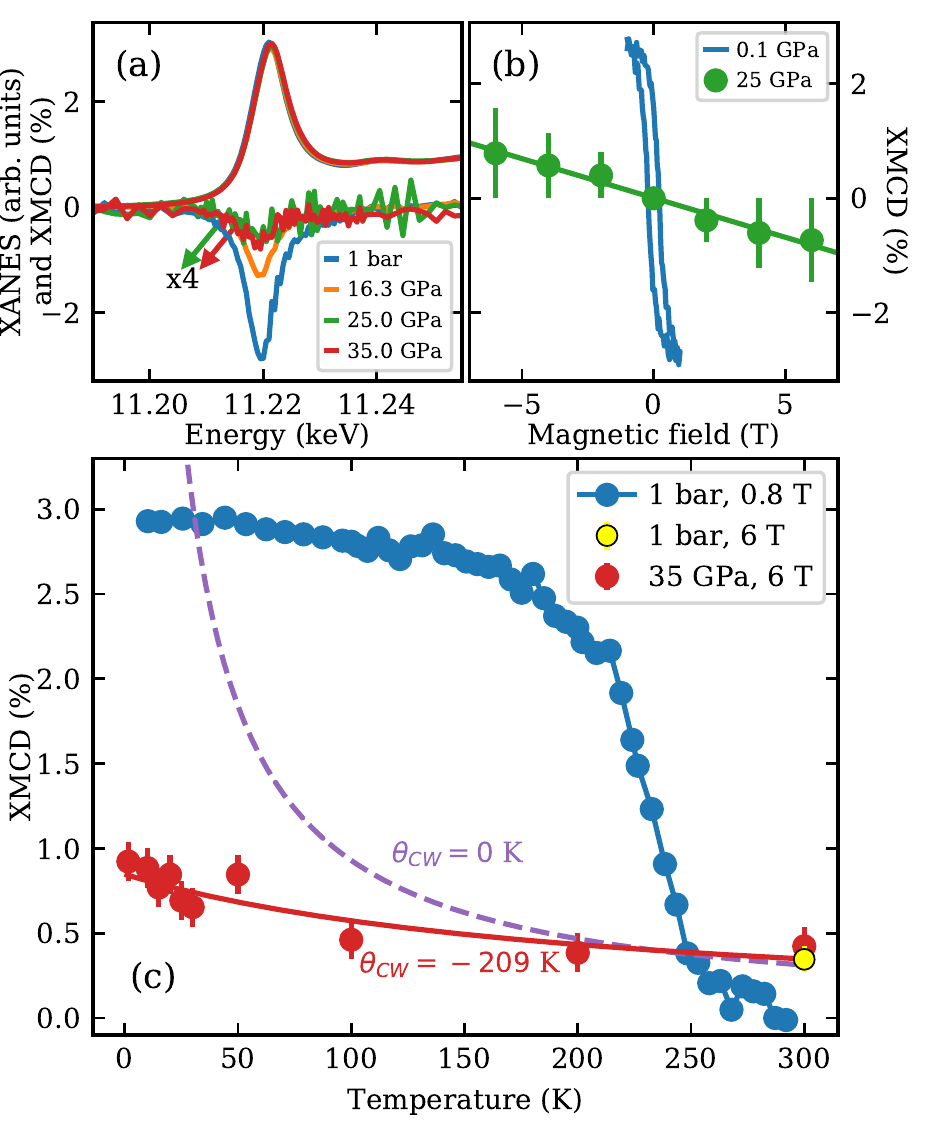}
\caption{(a) Ir L$_3$ x-ray absorption and XMCD spectra at selected pressures (H=1 T, T=1.6 K). The XMCD signal at 25 GPa and 35 GPa are quadrupled for clarity. (b) Field-dependent XMCD at 0.1 GPa (T=6 K) and 25 GPa (T=1.6 K) (c) Temperature dependence of XMCD signal in the ordered (1 bar) and disordered (35 GPa) magnetic phases. A Curie-Weiss model yields $\theta_{CW}$=-209 K for the disordered phase (see supplemental material for additional details). The susceptibility of an uncorrelated (Curie) paramagnet is shown with dashed lines. In a 6 T field, the susceptibility of the thermally disordered phase at ambient pressure (T=300K $>$ T$_N$=240 K) is similar to that of the quantum disordered phase.}
\label{fig3}
\end{figure}

Having established the absence of magnetic order in Sr-214 at high pressures, we now provide evidence in support of a pressure-induced magnetically disordered phase driven by in-plane quantum fluctuations (i.e., a quantum paramagnet). To this end we studied the magnetic field (up to 6 T) and temperature dependence of the Ir $L_3$ XMCD at 25 and 35 GPa. Despite a substantial reduction in XMCD intensity beyond 17 GPa, a small signal is observed in large magnetic fields [Fig.~\ref{fig3}(a)]. Noticeably, the room temperature magnetic response is pressure-independent up to at least 35 GPa [Fig.~\ref{fig3}(c)], demonstrating that the local Ir magnetic moment is preserved in this pressure range. Combined with the persistent insulating state and nearly constant $\langle L.S \rangle$ \cite{Haskel-PRL,Zocco} this result establishes that Sr-214 remains a $\mathrm{J_{eff}} = \frac{1}{2}$-like Mott insulator at least up to 35 GPa.

The temperature dependence of the XMCD intensity shows absence of magnetic order down to T=1.6 K even in a H=6 T applied field [Fig.~\ref{fig3}(c)]. As discussed above, if the in-plane exchange interactions were undisturbed, such large field would have induced magnetic ordering and a sizable W-FM response comparable to the response at ambient pressure (a 3.0\% XMCD signal at the Ir $L_3$ edge corresponds to 0.05 $\mu_B$/Ir \cite{Haskel-PRL}). A fit to a Curie-Weiss law yields $\theta_{\rm CW}=-209(40)$ K and an effective moment of $1.45(7)\mu_B$ \cite{Supplemental}. The magnitude of the effective moment is close to the theoretical value for $J=\frac{1}{2}$ Ir$^{4+}$ ions in the strong SOC limit \cite{Cava-PRB}. The large negative value of $\theta_{\rm CW}$ indicates strong local AFM correlations and a high degree of frustration in the intralayer coupling between localized Ir moments even at T=1.6 K [$f = |\theta_{\rm CW}|/T_{\rm min}\sim 140$ \cite{Balents-QSL}]. Note that a Curie paramagnet with no local exchange correlations would have resulted in $\times$15-20 larger XMCD signal at 6 T, 1.6 K (local Ir moment is in the $\sim 0.25-0.35~\mu_B$ range as determined by neutron scattering measurements \cite{Ye}). Although the strength of local exchange interactions is comparable to that in the thermally induced paramagnetic state of Sr$_2$IrO$_4$ at ambient pressure, $\theta_{\rm CW}=+236$ K \cite{Chikara-PRB-CW}, the sign of local exchange interactions is opposite between these two phases pointing to their distinct nature. The disordered magnetic state at high pressure is also distinct from typical disordered states obtained by doping the IrO$_2$ layers. For example, 3\% Tb doping destroys magnetic order but yields a negligible $\theta_{\rm CW}=-1.5$ K \cite{Cao-Tb}. The dramatic degree of magnetic frustration in the presence of strong exchange interactions suggests that Sr-214 is a quantum paramagnet above 17-20 GPa.

The field-dependent XMCD signal in the high-pressure phase is displayed in Fig.~\ref{fig3}(b). Both remanence and coercivity collapse, and a linear field dependence with no obvious discontinuities is observed. In general the latter would suggest an ungapped quantum paramagnet, e.g, as a result of D-M interactions rooted in spin-orbit coupling mixing singlet and triplet states. However, while the full spin Hamiltonian of Sr-214 at ambient pressure can be mapped to SU(2) symmetry by a local rotation \cite{Aharony, Senthil, Liu-NatPhys}, the coupling of magnetic moments to an external magnetic field involves a non-trivial $g$-tensor so that a measurement of the uniform spin susceptibility maps to a measurement of a linear combination of Q=0 and Q=($\pi, \pi$) susceptibilities \cite{Senthil}. The non-zero Q=($\pi,\pi$) component would give a non-zero magnetic susceptibility even in the presence of a gap so our data cannot rule out a gapped quantum paramagnet.

We now discuss possible routes to quantum paramagnetism in compressed Sr$_2$IrO$_4$. At ambient pressure, the dispersive magnetic excitations of Sr-214 have been measured by RIXS \cite{Kim-magnons} and fit to an isotropic Heisenberg model of interacting isospins with exchange constants $J_1 = 60$ meV, $J_2 = -20$ meV, and $J_3 = 15$ meV. This fit neglects ring exchange $J_r$ \cite{Coldea2001}, which, like a negative $J_2$, results in downward dispersion from Q=($\pi$,0) to Q=($\frac{\pi}{2},\frac{\pi}{2}$). Hence the values of $J_2$, $J_3$ and $J_r$ are not unequivocally known and could provide a path to frustration. Nevertheless, taken at face value, the reported exchange constants at ambient pressure correctly predict a N\'eel phase within the $J_1-J_2-J_3$ model \cite{Danu2016}, and place Sr-214 in close proximity to a classical critical phase boundary $J_3=0.5|J_2-0.5J_1|$ near which quantum fluctuations can drive quantum paramagnetism  \cite{Capriotti2004b}. It is thus plausible to conclude that a pressure-driven increase in orbital overlap and exchange interactions \cite{Overlap-note} pushes Sr-214 towards this phase boundary and quantum paramagnetism with predictions including columnar and plaquette valence bond solids (VBS) and spin liquid phases \cite{liquid2,Mambrini2006}. Note that the columnar VBS breaks in-plane lattice translational symmetry, an effect not detected within the accuracy of previous powder \cite{Haskel-PRL,Granado-XRD} or current single crystal diffraction measurements. It remains to be seen whether realistic Hamiltonians for exchange correlations between J$_{\rm eff}=\frac{1}{2}$ moments at high pressure can stabilize a quantum paramagnetic phase. In particular, the role of $J_r$ in Sr-214 needs to be experimentally and theoretically clarified as such exchange may also lead to quantum criticality \cite{Misumi2014,Larsen2019}. Measuring the spectrum of magnetic excitations of Sr-214 at high pressure remains critical; not only to determine changes in the exchange constants, but also to verify if the isotropic Heisenberg model remains a valid approximation.

In conclusion, we show a remarkable response of magnetic interactions in Sr$_2$IrO$_4$ to compression. Pressures of a few GPa alter the interplanar magnetic exchange and drive a magnetic crossover which mimics that observed under a magnetic field at ambient pressure \cite{Kim-Science}. More importantly, higher pressures modify the intraplanar exchange between $\mathrm{J_{eff}}=\frac{1}{2}$ isospins leading to a highly frustrated magnetic state ($f \sim 140$) and possible emergence of quantum paramagnetism. This frustration is likely driven by a pressure-induced enhancement of $J_2$, $J_3$, and/or $J_r$ exchange interactions relative to $J_1$, emphasizing the importance of exchange pathways beyond first neighbors in square lattices with extended $5d$ orbitals. Besides opening exciting questions on the detailed nature of the magnetic state of Sr-214 at high pressures, these results raise the prospect of tuning other $5d$ based systems into emergent phases. For instance, a recent work in the double-layer Sr$_3$Ir$_2$O$_7$ has provided evidence for a magnetic phase transition around 15 GPa, above which the system is argued to be in a frustrated paramagnetic state \cite{Zhang2019}. Finally, advances in experimental techniques, such as RIXS \cite{RIXS-JSR,RIXS-HPR} and nuclear resonant scattering \cite{NFS-SR}, are required to probe magnetic excitations and short-range spin correlations in iridates at pressures of tens of GPa and beyond in order to provide deeper insight into the nature of quantum paramagnetic phases.

\begin{acknowledgments}
 Work at Argonne was supported by the U.S. DOE Office of Science, Office of Basic Energy Sciences, under Contract No. DE-AC02-06CH11357. We thank GSE-CARS for use of their DAC gas loading facility \cite{Rivers2008}. GC acknowledges NSF support via Grants DMR 1712101 and 1903888. BJ Kim was supported by IBS-R014-A2.
\end{acknowledgments}

\bibliography{refs}

\begin{thebibliography}{77}
\bibitem [{Ove()}]{Overlap-note}%
  \BibitemOpen
  \href@noop {} {}\bibinfo {note} {Evidence for enhanced overlap of 5d orbitals
  under pressure comes from electrical resistance measurements which show a
  continuous reduction of the Mott gap to at least 20 GPa
  \cite{Haskel-PRL,Zocco}.}\BibitemShut {Stop}%
\end{thebibliography}

\end{document}